\documentclass[12pt]{article}

\usepackage{amsmath}
\usepackage{amssymb}
\usepackage[dvips]{graphicx}
\usepackage{cite}

\makeatletter
 
  \@addtoreset{equation}{section}
 \makeatother

\newcommand {\n}{\nonumber \\}
\newcommand {\tr}{\mbox{tr}}

\begin{document}
\setlength{\oddsidemargin}{0cm}
\setlength{\baselineskip}{7mm}

\begin{titlepage}
\begin{normalsize}
\begin{flushright}
\begin{tabular}{l}
\end{tabular}
\end{flushright}
  \end{normalsize}

~~\\

\vspace*{0cm}
    \begin{Large}
       \begin{center}
         {Covariant Formulation of M-Theory}
       \end{center}
    \end{Large}
\vspace{1cm}

\begin{center}
           Matsuo S{\sc ato}\footnote
           {
e-mail address : msato@cc.hirosaki-u.ac.jp}\\
      \vspace{1cm}
       
         {\it Department of Natural Science, Faculty of Education, Hirosaki University\\ 
 Bunkyo-cho 1, Hirosaki, Aomori 036-8560, Japan}

\end{center}

\hspace{5cm}

\begin{abstract}
\noindent

We propose the bosonic part of an action that defines M-theory. It possesses manifest SO(1, 10) symmetry and constructed based on the Lorentzian 3-algebra associated with U(N) Lie algebra. From our action, we derive the bosonic sector of BFSS matrix theory and IIB matrix model in the naive large N limit by taking appropriate vacua. We also discuss an interaction with fermions.

\end{abstract}

\vfill
\end{titlepage}
\vfil\eject

\setcounter{footnote}{0}

\section{Introduction}
\setcounter{equation}{0}

M-theory is strongly believed to define string theory non-perturbatively. Although BFSS matrix theory \cite{BFSS} and IIB matrix model \cite{IKKT} describe some non-perturbative aspects of string theory, covariant dynamics have not been derived, such as the covariant membrane action or a longitudinal momentum transfer of D0 branes. This is because $SO(1, 10)$ symmetry is not manifest in these models. 

BFSS matrix theory and IIB matrix model can be obtained by the matrix regularization of the Poisson brackets of the light-cone membrane theory \cite{deWHN} and of Green-Schwarz string theory in Schild gauge \cite{IKKT}, respectively. Because the regularization replaces a two-dimensional integral over a world volume by a trace over matrices, BFSS matrix theory and IIB matrix model are one-dimensional and zero-dimensional field theories, respectively. On the other hand, the bosonic part of the membrane action can be written covariantly in terms of Nambu bracket as $T_{M2} \int d^3 \sigma \sqrt{ \{ X^L, X^M, X^N \}^2}$, where L, M, N run $0, 1, \cdots, 10$
\footnote{This is also equivalent to Schild-type action: $T_{M2} \int d^3 \sigma \sqrt{g}\left(-\frac{1}{12}(\frac{1}{\sqrt{g}}\{ X^L, X^M, X^N \})^2+\Lambda\right)$.}
\cite{Nambu}. One way to obtain a SO(1, 10) manifest field theory is to regularize the Nambu bracket $\{ X^L, X^M, X^N \}$ \cite{Yoneya} by 3-algebra $[X^L, X^M, X^N]$ \cite{BLG1,Gustavsson,BLG2}\footnote
{
A formulation of M-theory by a cubic matrix action was proposed by Smolin \cite{Smolin1, Smolin2, Azuma} }. In this case, we obtain a zero-dimensional field theory. In order to reproduce BFSS matrix theory and IIB matrix model, the 3-algebra needs to associate with ordinary Lie algebra. Recently, the authors of \cite{Lorentz0, Lorentz1, Lorentz2, Lorentz3, Iso} found that such 3-algebra needs to have a metric with an indefinite signature. 

In this paper, we propose the bosonic part of an action that defines M-theory:
\begin{equation}
S=-\frac{1}{12}<[X^L, X^M, X^N]^2>,
\end{equation}
where $X^L$ are spanned by elements of Lorentzian 3-algebra associated with U(N) Lie algebra. This action defines a zero-dimensional field theory and possesses manifest SO(1, 10) symmetry. By expanding fields around appropriate vacua, we derive the bosonic sector of BFSS matrix theory and IIB matrix model in the naive large N limit. We discuss the fermionic part of the action of M-theory in the appendix.

\vspace{1cm}

\section{Action of M-theory}
\setcounter{equation}{0}

We propose a following action that defines the bosonic sector of M-theory,
\begin{equation}
S=-\frac{1}{12}<[X^L, X^M, X^N]^2>, 
\end{equation}
where $L, M, N = 0, 1, \cdots, 10.$ This action defines a zero-dimensional field theory and possesses manifest SO(1,10) symmetry. There is no coupling constant.

 The bosonic fields $X^M$ are spanned by elements of the Lorentzian 3-algebra,
\begin{equation}
X^M=X^M_{-1} T^{-1} +X^M_0 T^0 +X^M_i T^i,
\end{equation}
where $i=1, 2, \cdots, N^2.$
The algebra is defined by 
\begin{eqnarray}
&&[T^{-1}, T^a, T^b]=0, \n
&&[T^0, T^i, T^j]=[T^i, T^j]=f^{ij}_{\quad k} T^k, \n
&&[T^i, T^j, T^k] = f^{ijk} T^{-1},
\end{eqnarray}
where $a,b=-1,0,1,2, \cdots, N^2$ and $f^{ijk} = f^{ij}_{\quad l} h^{lk}$ is totally anti-symmetrized. $[T^i, T^j]=f^{ij}_{\quad k} T^k$ is the U(N) Lie algebra. The gauge symmetry of this theory is $N^2$-dimensional translation $\times$ U(N) symmetry \cite{Lorentz1}.    
The metric of the elements is defined by
\begin{eqnarray}
&& <T^{-1}, T^{-1}>=0, \quad <T^{-1}, T^{0}>=-1, \quad <T^{-1}, T^{i}>=0, \n
&& <T^{0}, T^{0}>=0, \quad <T^{0}, T^{i}>=0, \quad <T^{i}, T^{j}>=h^{ij}.
\end{eqnarray}

By using these relations, the action is rewritten as 
\begin{equation}
S= \tr (-\frac{1}{4}(X_0^L)^2 [X_M, X_N]^2 +\frac{1}{2} (X_{0}^{M}[X_M, X_N])^2),\label{rewritten}
\end{equation}
where $X^M=X^M_i T^i$ in the Lie brackets. There should be no ghost in the theory, because $X_{-1}^M$ do not appear in the action\footnote
{
Ghost-free Lorentzian 3-algebra theories were studied in \cite{ghost1, ghost2}.
}.

\vspace{1cm}

\section{BFSS Matrix Theory and IIB Matrix Model from M-Theory}
\setcounter{equation}{0}
Our theory possesses a large classical moduli that includes simultaneously diagonalizable configurations $[X^M, X^N]=0$ and arbitrary $X^M_0$. By treating appropriate configurations as backgrounds, we derive the bosonic sector of BFSS matrix theory and IIB matrix model in the large N limit. 
  
We consider backgrounds
\begin{equation}
\bar{X}_{\mu}=p_{\mu}=\mbox{diag}(p_{\mu}^1, p_{\mu}^2, \cdots, p_{\mu}^N), 
\label{bg1}
\end{equation}
where $\mu=0, 1, \cdots, d-1 (d \le 10)$.
$(p_{0}^i, p_{1}^i, \cdots, p_{d-1}^i)$ ($i=1, \cdots, N$) represent N points randomly distributed in a d-dimensional space. There are infinitely many such configurations. $X_0^M$ represents an eleven-dimensional constant vector. By using SO(1,10) symmetry, we can choose 
\begin{equation}
\bar{X}_0^M= \frac{1}{g} \delta^M_{10},
\label{bg2}
\end{equation}
as a background without loss of generality. $g$ will be identified with a coupling constant.

We expand the fields around the backgrounds, 
\begin{equation}
X_{\mu}=p_{\mu}+a_{\mu}. 
\label{expansion}
\end{equation} 
We do not integrate $X_0^M$ or the diagonal elements of $a_{\mu}$. Such procedure is known as quenching in the context of the large N reduced models. In the naive large N limit, if the full theory possesses supersymmetry, all the backgrounds (\ref{bg1}) and (\ref{bg2}) should be independent vacua and fixed without quenching \cite{KS}, as in the discussion of Higgs mechanism.

The first term of the action (\ref{rewritten}) is rewritten as
\begin{eqnarray}
S_1&=&\tr (-\frac{1}{4}(X_0^L)^2 [X_M, X_N]^2) \n
   &=&-\frac{1}{4g^2}\tr([p_{\mu}+a_{\mu}, p_{\nu}+a_{\nu}]^2 
+ 2[p_{\mu}+a_{\mu}, X^I]^2 
+ [X^I, X^J]^2),  
\end{eqnarray}
where $I, J, K = d, \cdots, 10$.
The second term is  
\begin{eqnarray}
S_2&=&\frac{1}{2} \tr((X_{0}^{M}[X_M, X_N])^2) \n
   &=&\frac{1}{2g^2} \tr([p_{\mu}+a_{\mu}, X^{10}]^2 + [X^{10}, X^I]^2).  
\end{eqnarray}
As a result, the total action is independent of $X^{10}$ as follows,
\begin{equation}
S=-\frac{1}{4g^2}\tr([p_{\mu}+a_{\mu}, p_{\nu}+a_{\nu}]^2 
+ 2[p_{\mu}+a_{\mu}, X^i]^2 
+ [X^i, X^j]^2), 
\end{equation}
where $i,j = d, \cdots, 9$.
In the large N limit, this action is equivalent to
\begin{equation}
S=-\frac{1}{4g^2} \int d^d \sigma \tr(F_{\mu\nu}^2 
- 2(D_{\mu} X^i)^2 
+ [X^i, X^j]^2). 
\end{equation}
This fact can be proved perturbatively and non-perturbatively \cite{EK,Parisi,BHN,GK}.

Therefore, if we choose the backgrounds with $d=1$, we obtain 
BFSS matrix theory in the large N limit,
\begin{equation}
S=\frac{1}{4g^2} \int d \tau \tr(
2(D_{0} X^i)^2 
- [X^i, X^j]^2). 
\end{equation}
If we choose those with $d=0$ (i.e. just choose $X_0^M= \frac{1}{g} \delta^M_{10}$), we obtain
IIB matrix model in the large N limit,
\begin{equation}
S=-\frac{1}{4g^2}\tr(
[X^i, X^j]^2). 
\end{equation}
We also obtain matrix string theory \cite{Motl, BS, DVV} when $d=2$.

\vspace{1cm}

\section{Conclusion and Discussion}
\setcounter{equation}{0}

In this paper, we have proposed a covariant action that defines the bosonic sector of M-theory. From this action we have derived the bosonic sector of BFSS matrix theory and IIB matrix model in the large N limit. By using these relations, we can directly discuss  covariant dynamics that have not been derived, such as covariant membrane and M5-brane actions and a longitudinal momentum transfer of D0 particles.

We discuss the reason why our theory with suitable backgrounds (\ref{bg1}), (\ref{bg2}) with $d=0$ and $d=1$ reduces to IIB matrix model and BFSS matrix theory, respectively, in terms of symmetry. First, we discuss the manner to reproduce IIB matrix model. If we compactify M-theory on a circle, we obtain type IIA superstring theory. The compactification radius is given by $R_{11}=g \sqrt{\alpha'}$, where $g$ is the string coupling constant and $\alpha'$ is an inverse of the string tension. Type IIA and IIB superstrings on circles are related by T-duality, where the radii satisfy  $R_{IIA} R_{IIB} = \alpha'$. Thus, the strong coupling limit of type IIB superstring with SO(1,9) Lorentz symmetry is realized by taking $R_{11} \to \infty$ and $R_{IIA} \to 0$ in M-theory with a torus compactification. The remaining Lorentz symmetry of M-theory is SO(1,9). Because the background (\ref{bg1}), (\ref{bg2}) with $d=0$ keeps this symmetry, our theory around it reproduces IIB matrix model. Next, BFSS matrix model is conjectured to describe the infinite momentum frame (IMF) limit of M-theory. In this limit, the remaining Lorentz symmetry is SO(9).  Because the background (\ref{bg1}), (\ref{bg2}) with $d=1$ keeps this symmetry, the theory around it reduces to BFSS matrix model.

We discuss a full action of M-theory in the appendix. We need to check whether it possesses supersymmetry. We are going to report the fermionic part in the next publication. 

As a first step, we have shown that our theory reproduces large N dynamics of the matrix models. In this limit, only planer diagrams contribute. As a second step, we should examine whether our theory includes all dynamics of the matrix models in order to check that it is a complete action of M-theory.

\vspace*{1cm}

\section*{Appendix}
\setcounter{equation}{0}
We consider a following action,
\begin{equation}
S=-\frac{1}{12}<[X^L, X^M, X^N]^2> 
+ \frac{1}{4}<\bar{\Psi} \Gamma_{MN}[X^M, X^N, \Psi]>, 
\label{fermionic}
\end{equation}
where $\Psi$ is a Majorana spinor of SO(1, 10). This action possesses manifest SO(1, 10) symmetry. It can be rewritten as
\begin{eqnarray}
S&=& \tr (-\frac{1}{4}(X_0^L)^2 [X_M, X_N]^2 
+\frac{1}{2} (X_{0}^{M}[X_M, X_N])^2 \n
&&\quad+\frac{1}{2}X^M_0\bar{\Psi}\Gamma_{MN}[X^N, \Psi]
-\frac{1}{2}\bar{\Psi}_0\Gamma_{MN}\Psi[X^M, X^N]).
\end{eqnarray}
If we expand the fields under the same conditions as in the bosonic case
(\ref{bg1}), (\ref{bg2}), and (\ref{expansion}) and the additional conditions
\begin{eqnarray}
&&\Psi_0=0, \\
&&\Gamma^{10}\Psi=\Psi,
\label{chiral}
\end{eqnarray}
we obtain
\begin{eqnarray}
S&=&-\frac{1}{g^2}\tr(\frac{1}{4}[p_{\mu}+a_{\mu}, p_{\nu}+a_{\nu}]^2 
+ \frac{1}{2}[p_{\mu}+a_{\mu}, X^i]^2 
+ \frac{1}{4}[X^i, X^j]^2 \n
&& \qquad + \frac{g}{2}\bar{\Psi}\Gamma^{\mu}[p_{\mu}+a_{\mu}, \Psi] 
+ \frac{g}{2}\bar{\Psi}\Gamma^{i}[X_i, \Psi]).
\end{eqnarray}
In the large N limit, this action is equivalent to
\begin{eqnarray}
S&=&-\frac{1}{g^2} \int d^d \sigma \tr( \frac{1}{4}F_{\mu\nu}^2 
- \frac{1}{2}(D_{\mu} X^i)^2 
+ \frac{1}{4}[X^i, X^j]^2 \n
&& \qquad \qquad \qquad
+ \frac{i}{2} \bar{\Psi} \Gamma^{\mu} D_{\mu} \Psi 
+ \frac{1}{2} \bar{\Psi} \Gamma^i [X_i, \Psi]),
\end{eqnarray}
where $\Psi$ is redefined to $\frac{1}{\sqrt{g}}\Psi$.
This is the maximally supersymmetric Yang-Mills theory in d-dimension. Thus, we obtain IIB matrix model, BFSS matrix theory and matrix string theory if we choose the backgrounds with d=0, 1, and 2, respectively. 

Therefore, (\ref{fermionic}) may be a complete action that defines M-theory. There are at least two things to examine. First, we need to check whether (\ref{fermionic}) possesses supersymmetry. Second, we impose the chirality condition (\ref{chiral}) by hand in the process to derive the matrix models. Therefore, we also need to check that the chirality condition (\ref{chiral}) is automatically satisfied in the large N limit. A planer diagram that includes fermions with opposite chirality may be forbidden.

\end{document}